%% file: samplepaper.tex
\documentclass[runningheads]{llncs}
\usepackage{graphicx}
\usepackage{amssymb}
\usepackage{wrapfig}
\begin{document}
\title{Tumera: Tutor of Photography Beginners}
%
%
\author{Xiaoran Wu \and
Jia Jia}
\authorrunning{X. Wu et al.}
%
\institute{Department of Computer Science and Technology, Tsinghua University, China \\
\email{846120867@qq.com}, \email{jjia@mail.tsinghua.edu.cn}}
%
\maketitle              
\begin{abstract}
With the popularity of photographic equipment, more and more people are starting to learn photography by themselves. Although they have easy access to photographic materials, it is uneasy to obtain professional feedback or guidance that can help them improve their photography skills. Therefore, we develop an intelligently interactive system, Tumera, that provides aesthetics guidance for photography beginners. When shooting, Tumera gives timely feedback on the pictures in the view port. After shooting, scores evaluating the aesthetic quality of different aspects of the photos and corresponding improvement suggestions are given. Tumera allows users to share, rank, discuss, and learn from their works and interaction with the system based on the scores and suggestions. In the experiment, Tumera showed good accuracy, real-time computing ability, and effective guiding performance.

\end{abstract}
\input{1-Introduction}
\input{2-Related}
\input{3-System}
\input{5-Experiments}
\input{6-Conclusion}
%
%
%
%

\end{document}

%% file: 1-Introduction.tex
\section{Introduction}

With the popularity of photographic equipment and picture-sharing apps stimulated by the rapid development of the Internet, people's enthusiasm for learning photography has gradually increased. Rather than strong artistic motivations and ambitious photography goals, people typically pursue direct visual appealingness so that their photography work can be more good-looking and be distinguished from the large number of photos shared by their friends. However, as the well-known American art critic Susan Sontag said in her series of essays "On Photography"~\cite{ref_sontag_on}, "an ugly or grotesque subject may be moving because it has been dignified by the attention of the photographer. A beautiful subject can be the object of rueful feelings, because it has aged or decayed or no longer exists." People's understanding of photography and what is beautiful can hardly be unified, and photography beginners need a certain amount of training before they form certain aesthetics concepts of photography.

With such a huge demand, various photography guide books~\cite{peterson2015learning,palangio2018practical} and courses are gaining popularity. Photography guide books are divided mainly into three categories: the use of equipment, post-processing~\cite{hu2018white,bie2019optimizing}, and aesthetic theory~\cite{beilin1991developmental,warren2002show,petursdottir2014imaging}. Although they provide guidance and support from different aspects, the non-interactive nature of books means the lack of feedback, preventing effective learning of beginners. On the other hand, photography courses make timely feedback possible, but they are expensive and demand photographic equipment, e.g., SLR. Moreover, guidance concentrates during classes. Consequently, students can not improve continuously under professional guidelines after courses.

Interactive photography guidance systems is an accessible extension of traditional photography guidance books and courses. Relying on a portable mobile terminal like a phone, these apps hold the promise to provide feedback to users quickly. However, most current photography assistance systems mainly focus on five functions: editing, filters, correction, pre-stage, and gallery. Editing apps finetune photos that have been taken, such as adjusting the brightness, saturation, and skin color level of the portrait. Filter apps serve similar aims and adjust the original image's appearance with the filters inherent in the app. Correction apps are mainly for local tine-tuning of photos, such as removing stains in the picture and simulating the effect of large aperture. Pre-stage apps adjust the parameters of the mobile phone camera and make the mobile phone a small SLR. The last kind, gallery apps, such as National Geographic and Mapworm, display excellent photography works but do not guide photography. These five types of photography assistance systems do not take advantage of mobile devices' flexible user interaction interfaces and largely lack timely feedback and guidance on shooting actions.

To enable users to learn photography skills more efficiently, in this paper, we develop an interactive photography guidance system, Tumera. We propose that the photograph learning process be divided into three steps -- perception, action, and feedback. Current photography apps directly process photos via pre-defined algorithms or show selected photos, ignoring  the actions of users. In contrast, we aim to build a system that provides feedback based on users' actions, thereby improving their aesthetic sense of perception and photography skills.

Specifically, Tumera achieves this goal by implementing a PAF (Perception, Action, and Feedback) interactive guidance system for photography beginners. The system contains the following components. First, at the perception level, our method meets users' need to get a calibration about aesthetic evaluations about perception. Tumera gives an overall aesthetic score and six feature scores assessing different aspects of the photo, i.e., balanced elements, color harmony, object emphasis, good lighting, rule of thirds, and vivid color. Secondly, at the action level, our system simplifies the shooting and upload process of photography beginners and provides real-time guidance on obtaining photos of higher quality. Thirdly, at the feedback layer, through a comprehensive analysis of the photographic works uploaded by the photography beginners, suggestions for improvement are given to help users understand the aesthetics of their works. 

Technically, this article uses a multi-task convolutional neural network (CNN) to train two scoring tasks at the same time. The main task is to predict the aesthetic score, and the second task is to predict scores of different aesthetic features. This neural network model mainly includes three parts: feature extraction, spatial pyramid pooling, and multi-task regression. A pre-trained ResNet extracts image features. The advantage is that we can adapt quickly to the aesthetics evaluation task at hand via training on a small dataset. To train a CNN model efficiently by images in batch, previous methods resize or crop images to have the same size, which is detrimental to the aesthetics evaluation task because resizing or cropping changes the original image layout, structure, and texture details. Tumera uses a spatial pyramid pool (SPP) algorithm and generates fixed-length representation vectors for images without resizing or cropping. 

The PAF interactive aesthetic evaluation system proposed in this paper not only opens up new application ideas for the research field of computer image aesthetics evaluation but also provides a practical basis for the application of photography guidance in leading the entire learning process of photography beginners. For evaluation, we invite 30 volunteers to participate. Each volunteer is asked to take a photo on a spot and upload it to the Tumera system. After that, according to its computational model, Tumera gives improvement suggestions, and the volunteers are asked to take another photo at the same location and upload it to Tumera again. We invited three professional photographers with different background to score the uploaded photos. Photographers choose the better ones from the two pictures taken by the volunteers. The effectiveness of our system is validated by the ratio of post-guidance photos in the photographer-selected set. In addition, we carry out experiments to demonstrate an alignment between scores given by Tumera and photography professionals, verifying the effectiveness of our aesthetics evaluation model. 

%% file: 2-Related.tex
\section{Related Works}

In this section, we introduce the development history of photography guidance and related works on photography aesthetic evaluation. 

\subsection{Photography guidance}
Early photography guidance relies on books. Experienced people summarize technical and aesthetic experiences about photography, such as design, color, and composition in photography through words and pictures~\cite{peterson2015learning,peterson2016understanding}. Traditional photography guide books can be traced back to the world’s first photography publication, "Photographs of British Algae: Cyanotype Impressions."~\cite{atkins1843photographs} This book has always been a collection of the New York Public Library since its first publication. The author of this book hand-processed and printed more than 6,000 photos in order to produce the books.

Later, due to the diversification of education and the incremental demand of the photography market, photography courses emerge. Some well-known art colleges and universities begin to offer photography courses, such as the world-renowned art and design college, Rhode Island School of Design, which aims to train students to explore the technology, concepts, and aesthetics of photography in the broad context of society and culture. At the same time, students will learn about the production, display, and understanding of images, and explore the different skills of photography. Moreover, colleges teach students how to use cultural symbols and metaphors to express their concerns through photography.

\subsection{Aesthetic Evaluation}
The traditional method of evaluating image aesthetics is mainly based on manually designed feature extractors. Initially, image aesthetics evaluation researchers use frame features to represent the aesthetic features of the image. The work of Datta et al.~\cite{datta2006studying} and Ke et al.~\cite{ke2006design} first transform the aesthetic understanding of images into a binary classification problem~\cite{dou2019webthetics,tian2015query,lu2015rating}.~\cite{datta2006studying} combines low-level and high-level features, usually used for image retrieval, and trains an SVM classifier~\cite{joachims1988making} for aesthetic image quality binary classification.~\cite{ke2006design} proposes to use color distribution, hue count, global edge distribution, and brightness as features to represent images and then forms a Bayesian naive classifier based on these features. These pioneering works use hand-designed features to model the global aesthetics of an image computationally.

Learning image features from a large amount of data has shown better performance in tasks such as recognition, retrieval, and tracking, surpassing the ability of traditional hand-designed features. Since Krizhevsky et al.~\cite{krizhevsky2012imagenet} used convolutional neural networks for image classification, more and more researchers have begun to learn image representations through deep learning methods.

To the best of our knowledge, current deep learning research mainly focuses on (a) How to design the network structure and network input, and retain the global information and local details of the image when the size of network input is fixed; (b) How to use the image style and semantic information, or how to choose an aesthetic quality evaluation model suitable for images with different content; (c) The format in which the aesthetic quality score is given, such as binary classification, regression, sorting, etc.

%% file: 3-System.tex
\begin{figure}[t!]
    \centering
    \includegraphics[width=\linewidth]{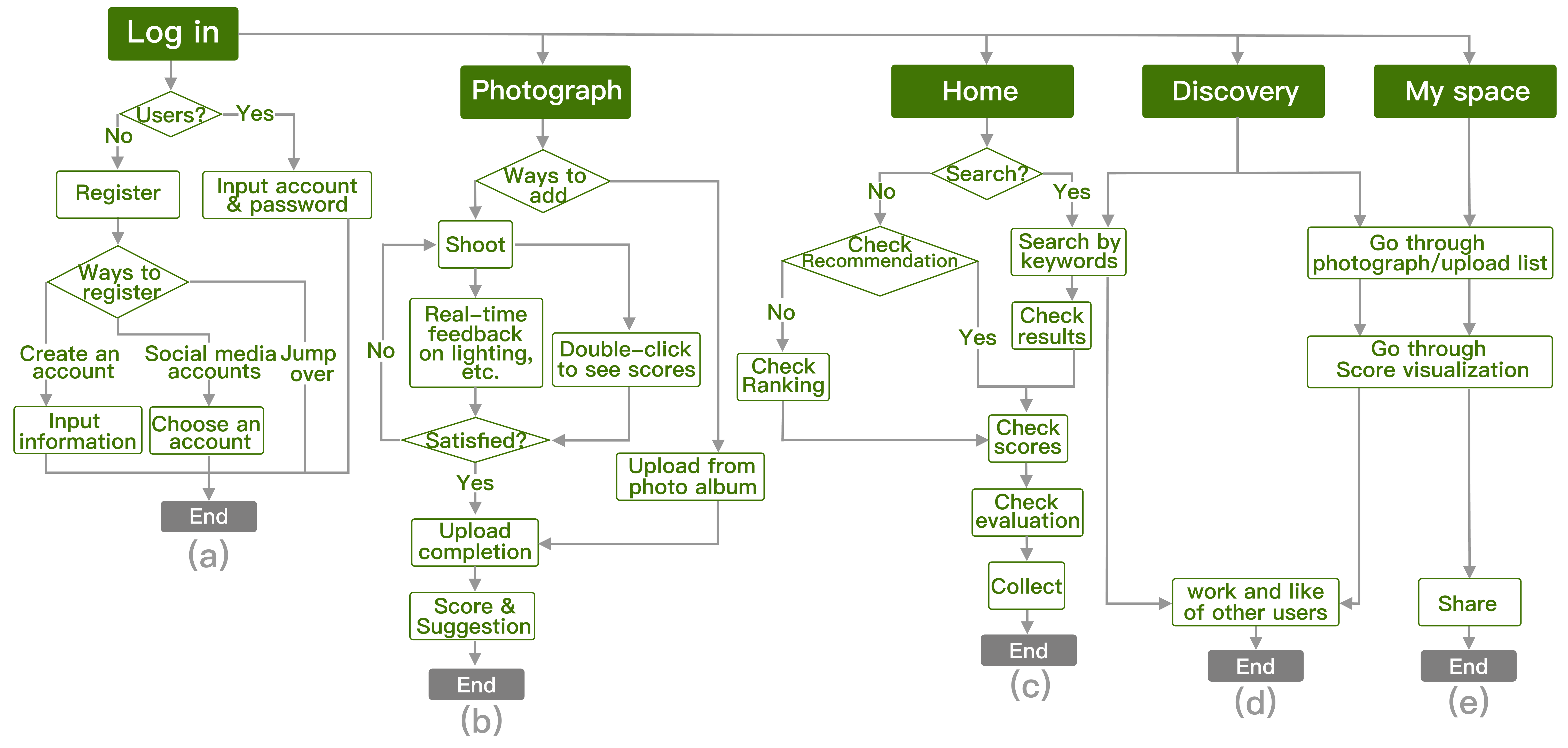}
    \caption{Scheme of the proposed Tumera system.}
    \label{fig:framework}
\end{figure}

\section{Tumera System}

In order to improve the learning efficiency of and cognitive feedback provided to photography beginners, enable photography beginners to better accumulate experience about photography aesthetics through guidance and feedback, and help photography beginners to shoot scenery with aesthetic value quickly, we design the Tumera system (Fig.~\ref{fig:framework}) that explores a variety of interactive methods to improve user experience.

In the design and implementation of the photography guidance application, the main consideration is how to guide users in photography activities effectively. To this end, we propose to model users' learning process as a perception-action-feedback (PAF) process. Users perceive their local observations, take (shooting) actions accordingly, and get feedback afterward. Feedback would influence how users perceive their environment and what actions users would take. 

\subsection{Overview}
\begin{wrapfigure}[19]{r}{0.5\linewidth}
    \vspace{-2em}
    \includegraphics[width=\linewidth]{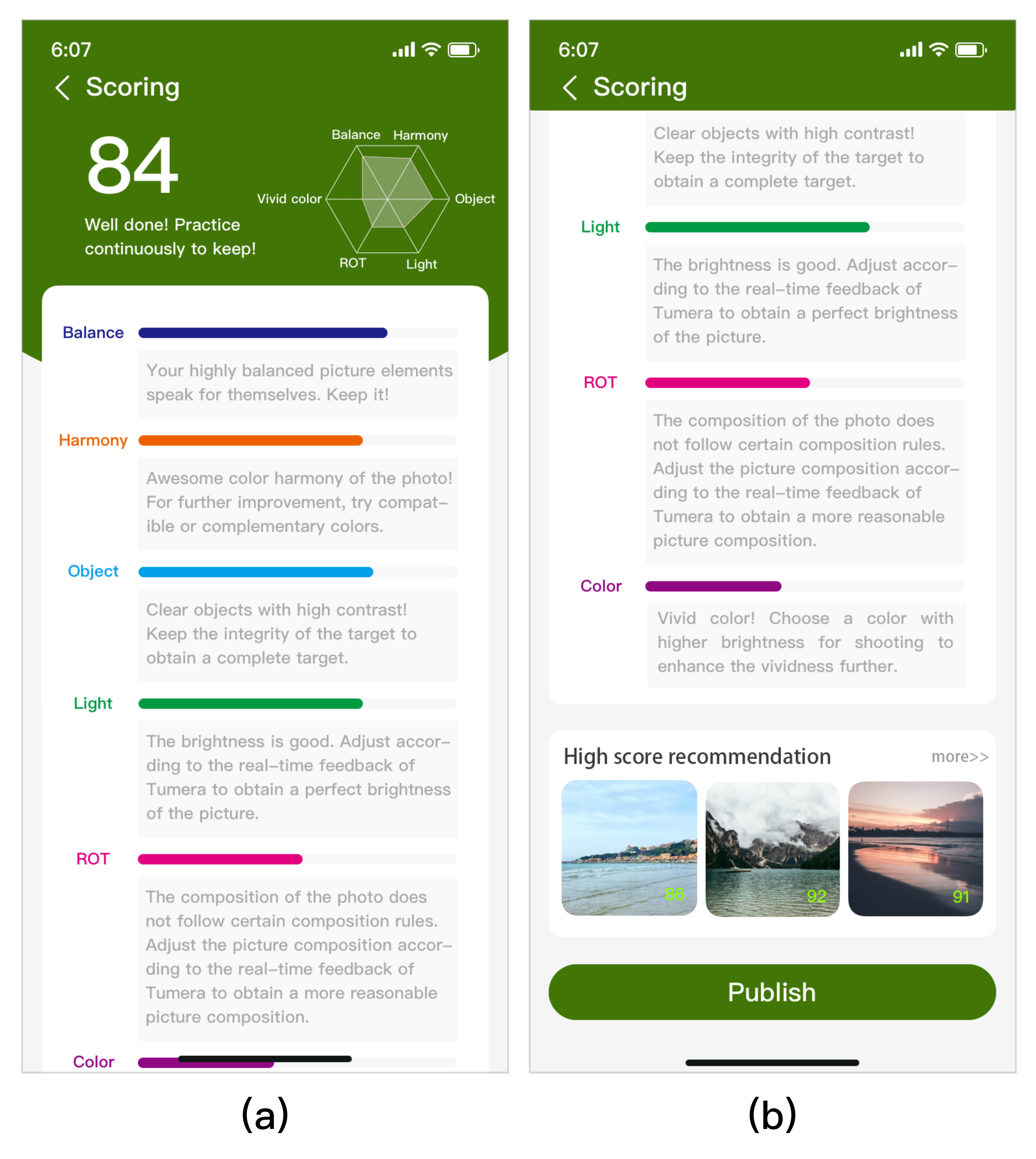}
    \caption{Tumera scoring interface.}
    \label{fig:score}
\end{wrapfigure}
Tumera is a photography assistant guiding all the three steps (perception, action, and feedback) of the PAF model tailored for the user. In this way, Tumera incorporates photography guidance and injects learning and feedback functions into the whole photography process. At the \emph{action} layer, the user’s photography skills are mainly prompted through the system's voice or text prompts for real-time guidance to help the user in the detailed photographic action (Fig.~\ref{fig:framework} (b)). At the \emph{feedback} level, users are given scores evaluating the 6 characteristic dimensions of the user's photographic works and a total score estimating the overall appearance of the photos (Fig.~\ref{fig:framework} (c)). At the \emph{perception} level, users can upload photos to the Tumera community, where they can see the scores given by the system, comments shared by other users, and high-quality images recommended by the system (Fig.~\ref{fig:framework} (d-e)). In this way, users accumulate aesthetics experience via interaction with the system and other people.

In the hope that any photographer can quickly get started, we simplify the operation process of Tumera. We layout our components one by one guided by the PAF model (Fig.~\ref{fig:framework}). Users first log in or create their accounts. After that, they can choose to take a photo under the real-time and interactive guidance of Tumera, search and compare high-quality photos, share photos with other photography enthusiasts. In the following sections, we introduce the interaction design (Sec.~\ref{sec:id}), data visualization (Sec.~\ref{sec:id}), and scoring scheme (Sec.~\ref{sec:ss}) of the Tumera system. 

\begin{wrapfigure}[21]{l}{0.5\linewidth}
    \vspace{-2em}
    \includegraphics[width=\linewidth]{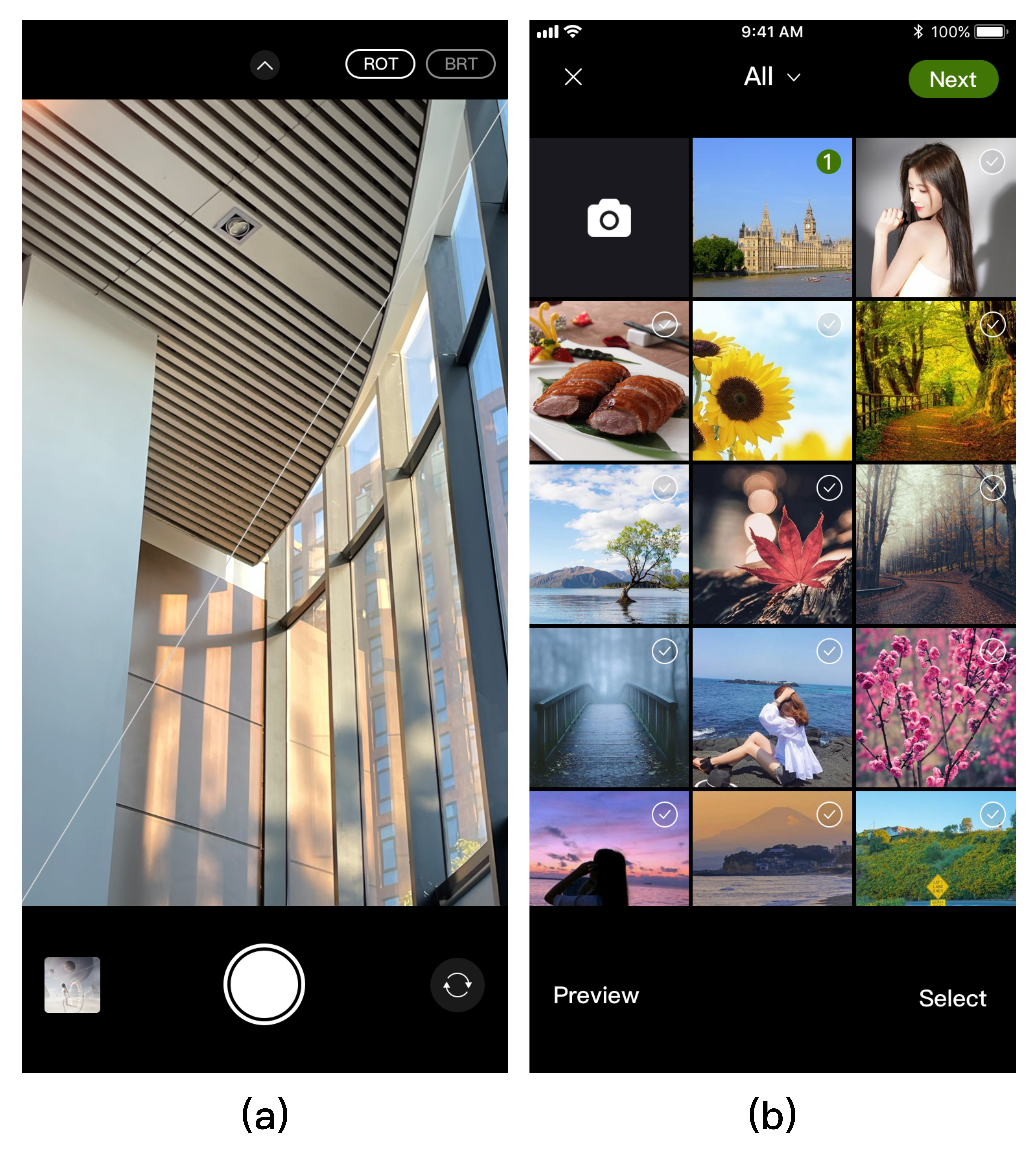}
    \caption{(a) Tumera guidance on picture configuration. (b) Upload a picture for aesthetic scoring.}
    \label{fig:shoot}
\end{wrapfigure}
\subsection{Tumera interaction design}\label{sec:id}
Real-time feedback during the shooting process is a critical component of Tumera in improving users' aesthetics concept and photography skills. To provide supportive, real-time, convenient, and comprehensive guidance for shooting, we include a variety of interaction methods in the application design: voice interaction, text reminder interaction, click interaction, and interaction among users. The interaction comes in the form of dynamic effects and user-friendly visual visualization to improve user experiences.

\textbf{Voice interaction.} During the shooting process, the user is prompted by the system for both picture composition and lighting conditions to obtain better photographs. Compared to text prompts, during shooting, voice prompts would not disturb the user's attention on the scenery picture when the user concentrates on the scenery in the viewport. For example, the voice interaction is used to provide information about the light. If the picture in the viewport is too bright, the system will give a reminder "too bright," and the system will also give a reminder like "too dark." Through voice interaction, Tumera is like a photography teacher in the classroom during the overall shooting process, giving the beginners of photography timely feedback on the light and composition and helping the beginners quickly build up the aesthetics of photography. Motivational sentences in voice are another feature of the Tumera system, such as "awesome," "yes," "good shot," and other positive voice feedback, ensure that users feel motivated during use.

\textbf{Click interaction.} Compared to others, some interactive operations are revoked less frequently and are more computationally expensive, such as real-time scoring of the picture in the viewport. For these operations, we request users to double-click the screen to start the interaction. Click interaction exists in the perception layer and feedback layer of the PAF conceptual model. In this way, the user can directly view the score and suggestion of the current picture in the viewport with a double-click operation. 

Regarding the design of click animation, the author starts from the visual design criteria, prioritizing showing clear information to the user. In addition, the author stresses bringing the user unified information guidance, and the prompts are in the same style as other prompts.

\begin{wrapfigure}[20]{r}{0.5\linewidth}
    \vspace{-2em}
    \includegraphics[width=\linewidth]{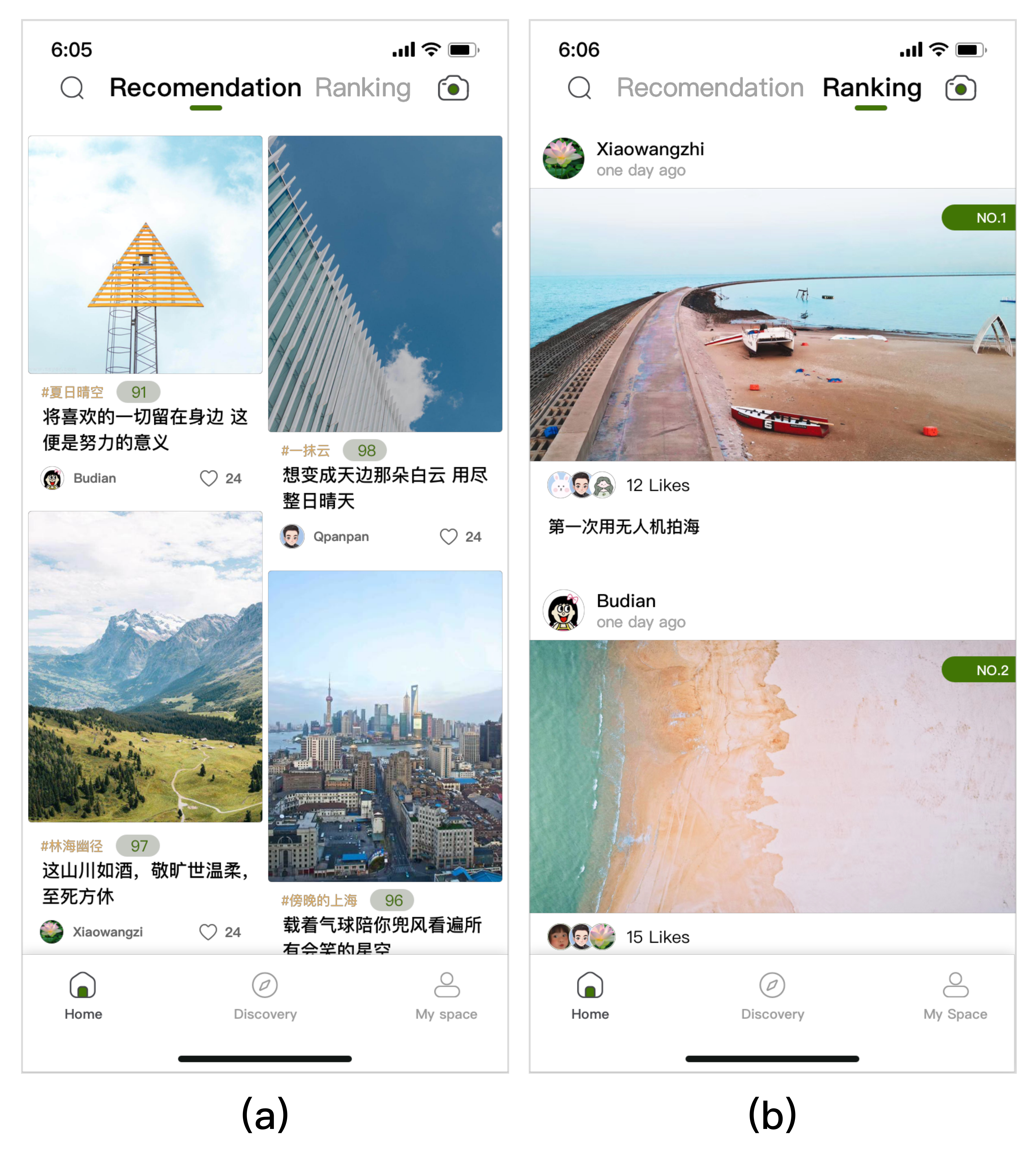}
    \caption{Tumera recommendation and ranking list.}
    \label{fig:recommend}
\end{wrapfigure}
\textbf{Real-time camera feedback interaction.} Sometimes, during shooting, suggestions need urgent user attention. In Tumera, the interactive mode for the prompts of composition is not confined to voice interaction but also text prompts on the interface to remind the user of the most suitable composition rule. These text prompts are combined with the user's operation, give prompts like "left", "right", "up", "down", "forward", and "backward". The composition guide in the Tumera system is based on several composition rules commonly adopted in photography, such as the rule of thirds, symmetrical composition, diagonal, framed composition, triangle, and center composition. We run an object detection algorithm to analyze the current photo composition. If it does not allow any composition rule, we prompt users to improve their photos in the view port.

Tumera's guidance is mainly used for landscape photography. Portrait photography also requires feedback about the overall construction of photos, light, and composition, so Tumera can also give feedback on these aspects when shooting objects are people, helping photographers to improve in multiple dimensions.

\textbf{Interaction among human users.} Peer motivation is believed to exert a significant influence on users' learning. In Tumera, we exploit this point and design interaction among users in multiple forms. 

First, Tumera provides users with recommendations of high-scoring photographic works. At the same time, there is also a ranking of photographic works uploaded in the community that day, which encourages users to create, share, comment, and learn through Tumera.

Moreover, after users get feedback, an incentive mechanism is established to promote user enthusiasm: users can share high-scoring photographic works through cards on social platforms. Users can also view past scoring records in the personal center. Excellent high-scoring photos will be shown on the homepage, inspiring users' enthusiasm for aesthetic creation.

As an application that emphasizes photography guidance but less social interaction, Tumera weakens the text when sharing, focusing on the inspiration for other users in the presentation of photographic works and scoring. Additionally, users can find relevant photography under different topics.

\textbf{Data visualization.} For the scoring results (Fig.~\ref{fig:score}), the presentation of data gives users an evaluation of the aesthetic calculation results. What is brought to the user is not just a number but also an emotional visual assessment. To this end, among the scores, we give priority to displaying the user’s highest score, no matter which one of the 6 dimensions. Scores are displayed in the form of running dynamic effects and displayed in a track. Encouraging prompts like "It's almost the end," "Persevere to improve," and other different sentences accompany the user's scoring.

\subsection{Scoring model}\label{sec:ss}
In this section, we introduce the computational model used by Tumera to score the photos.

\textbf{Using ResNet to extract features.} In implementing Tumera system, due to the high training cost and expensive manual labeling, we use a pre-trained ResNet~\cite{he2016deep} to extract image features. ResNet models trained on ImageNet have been proven to be sufficient for feature extraction. Therefore, although the data set we use, AADB~\cite{kong2016photo}, is too small to train accurate models with too many parameters, we can still obtain well-formed image representation.

Although the RGB color space is more convenient to display, the HSV color space~\cite{pang2011color} is closer to the real human vision. Therefore, it is undoubtedly more appropriate to adopt the HSV color space for the aesthetics evaluation of photos. However, the HSV color space also has shortcomings, that is, there is only one dimension for distinguishing colors, so the ability to express complex colors is limited. The LAB color space can solve this problem and may be a more suitable color space~\cite{murray2017deep}. Therefore, the LAB color space and HSV color space are compared and explored in this paper.

Our model uses a 7-dimensional output, containing scores for 6 aesthetic features and 1 overall score prediction. In the existing data set, the AADB data set~\cite{kong2016photo} meets the training requirements because it contains 10,000 pictures, and each picture has an overall score and a score for each of the aesthetic features. Currently, the largest data set of picture aesthetics is the AVA data set~\cite{murray2012large}. In this data set, there are a large number of baselines for reference. In order to better verify the network performance, a model that only calculates the overall aesthetic score is trained in the AVA data set.

For ordinary CNN networks, the superposition of layers can easily lead to gradient dispersion or gradient explosion, that is, the closer the CNN parameters to the input end, the more extreme the parameters will be during backpropagation. Parameters may be too small (gradient dispersion) or too large (gradient explosion). Both of these effects would seriously affect network performance. The proposal of ResNet solves this problem by using a residual structure.

In this paper, the AADB data set is input to ResNet to extract the features, the AVA data set is used as the test set, and the Spearman correlation coefficient and accuracy are used as evaluation indicators. The data ratio of the training set to the test set is 9:1.

\textbf{Spatial pyramid pooling.} For training a general image processing network, pictures are often scaled or intercepted, and subsequent processing calculations are performed after the pictures are converted to a specified size. However, for aesthetic recognition, these two methods are problematic. Non-proportional zoom will destroy the beauty of the picture, and the interception is likely to affect the original image configuration.

The CNN part of the network actually does not require a fixed size of the input photos. Large photos can be convolved and pooled to extract large feature maps, and small pictures correspond to small ones. Thus, the convolutional part does not require a fixed size of the input image. But when it comes to the regression and decision-making parts consisting of fully connected layers, it is necessary to turn feature maps of different sizes into vectors of a uniform length.

Spatial pyramid pooling is a dynamic pooling method. For input of variable size, 4*4 features are first extracted through the largest pool, and then 2*2 is extracted by transforming the size. The feature is finally 1*1. Therefore, for any size input, a fixed-size vector can be obtained through SPP pooling, which is then passed through the final fully connected layer.

\begin{wrapfigure}[22]{l}{0.5\linewidth}
    \vspace{-2em}
    \includegraphics[width=\linewidth]{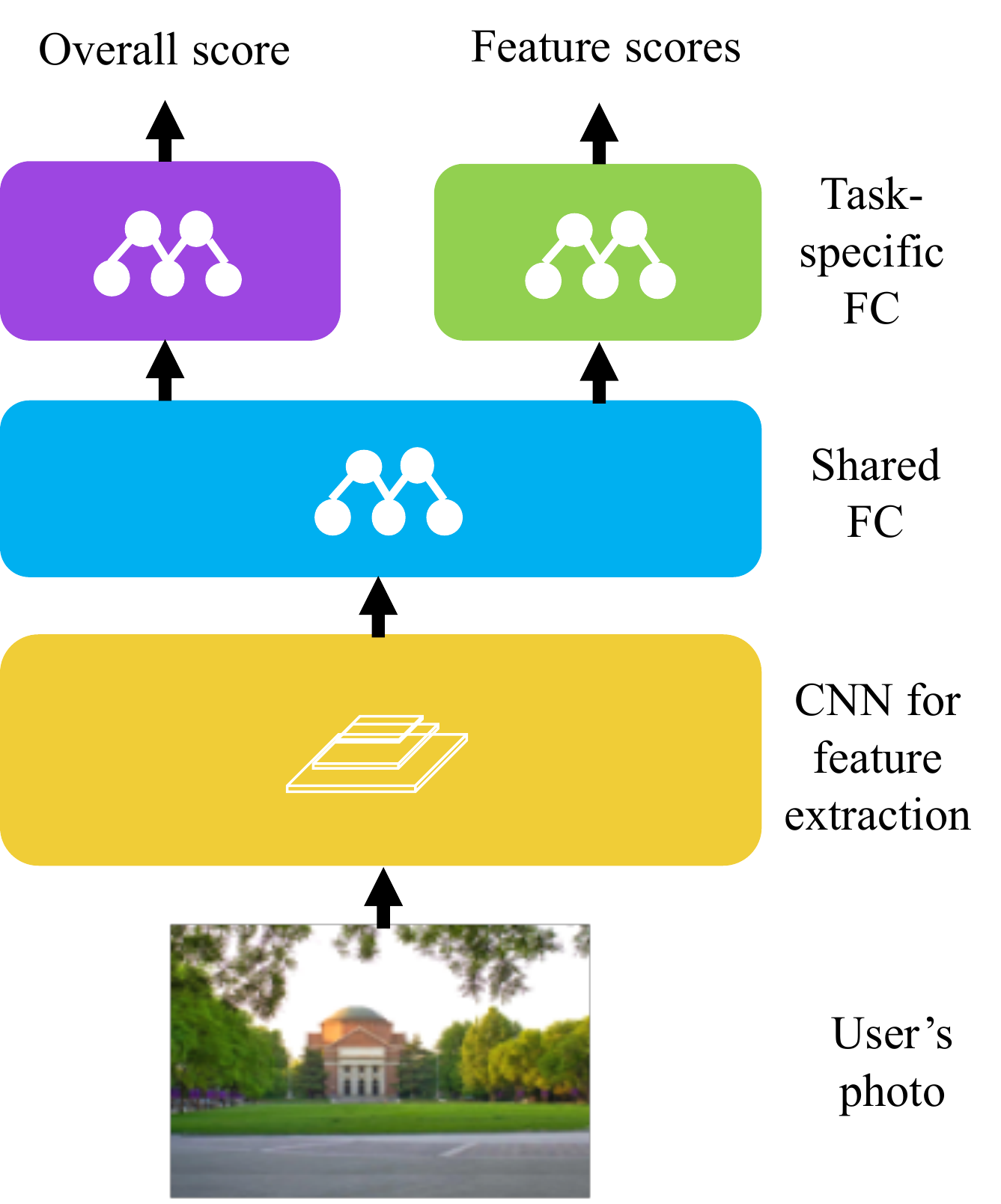}
    \caption{Network architecture for multi-task prediction. FC is short for fully-connected layer.}
    \label{fig:nn}
\end{wrapfigure}
\textbf{Multi-task regression decision}
Our network needs to learn two different tasks at the same time. One is the overall aesthetic score of the picture, and the other is scores of the picture on different aesthetic aspects (balanced elements, color harmony, object emphasis, good lighting, rule of thirds, and vivid color). These two tasks have independent parts, that is, the focus on different evaluation goals. In the meantime, they share common contents that are all manifestations of aesthetic properties. A network model that can exploit such a semi-sharing structure and can accomplish two tasks is expected.

The final structure adopted in this paper mimics the Multi-task cascaded convolutional neural networks (MTCNN)~\cite{kao2017deep}. In this model, different tasks share the same underlying feature extraction part, but split in the decision-making part. There are independent, fully connected layers, which respectively perform individual fitting tasks.

The final structure is shown in Fig.~\ref{fig:nn}. The yellow part is the feature extraction network, which is a pre-trained ResNet in practice. After this part is fully-connected networks for different tasks. It can be seen that these two tasks still share some of the fully connected layers (the blue part) before being separated into two independent judgment networks (the purple and green part).

The main training issue that needs attention is the setting of the loss function. We need to consider the balance of the two tasks in the loss function. In the final model, one task only needs to output the overall aesthetic core, while the other task needs to output the aesthetic score of each feature. The overall aesthetic score in the loss function must have a much higher weight than a single feature in order to ensure the balance of training.

%% file: 5-Experiments.tex
\section{Experiments}

Based on the proposed PAF photography guidance concept model for photographers, we compare the photography results before and after using Tumera real-time shooting guidance. By shooting the same scene, the scores obtained before using Tumera are compared with the scores of the pictures taken after the guidance of Tumera.

Three invited photography professionals: an advertising photographer, a photography graduate student, and a college photography teacher, each chooses the better one from photos taken before and after using Tumera. We then check whether the selected pictures are the ones with the guidance of Tumera. We also verify whether the evaluations of the Tumera model and photography professionals are consistent with each other.

\begin{wrapfigure}[14]{r}{0.5\linewidth}
    \vspace{-2em}
    \includegraphics[width=\linewidth]{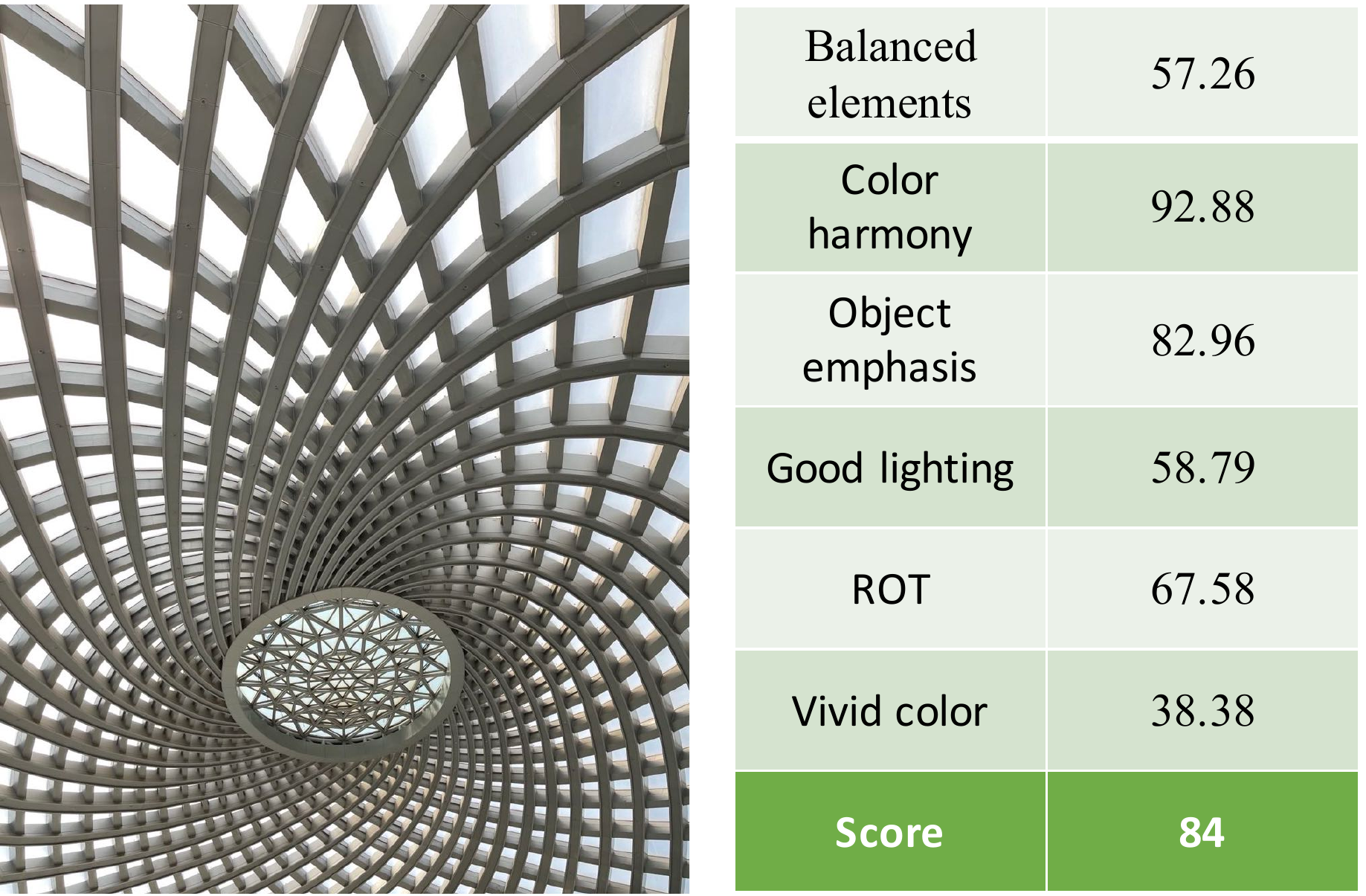}
    \caption{Tumera aesthetic scores for an example photo.}
    \label{fig:nn}
\end{wrapfigure}
\subsection{Experimenters}
In this experiment, we randomly recruit volunteer experimenters, regardless of age or gender. Before conducting the experiment, we confirm that every participant is a beginner in photography to ensure that our aim of helping beginners is being validated. We divide all experimenters into three groups and ask them to shoot at three different time periods (morning, noon to afternoon, afternoon to evening) to eliminate the external influence of light.

In order to ensure the fairness of our experiments, we select three photography professionals with different background, and subjectively judged the photographic aesthetic quality of photos from the perspectives of commercial photography, academic photography, and photography research.

\subsection{Implementation process}
\emph{Comparison of two photos for beginners.} We randomly recruited 30 photography beginners at Tsinghua University and asked them to shoot with the mobile phone camera in a certain scene and upload them to the Tumera system. After checking scores and suggestions provided by Tumera, volunteers are asked to shoot at the same scene for the second time with real-time guidance. Photos are uploaded again to the Tumera system. Table~\ref{tab:diff} shows the scores of photos shot before and after using Tumera.

\begin{figure}
    \centering
    \includegraphics[width=\linewidth]{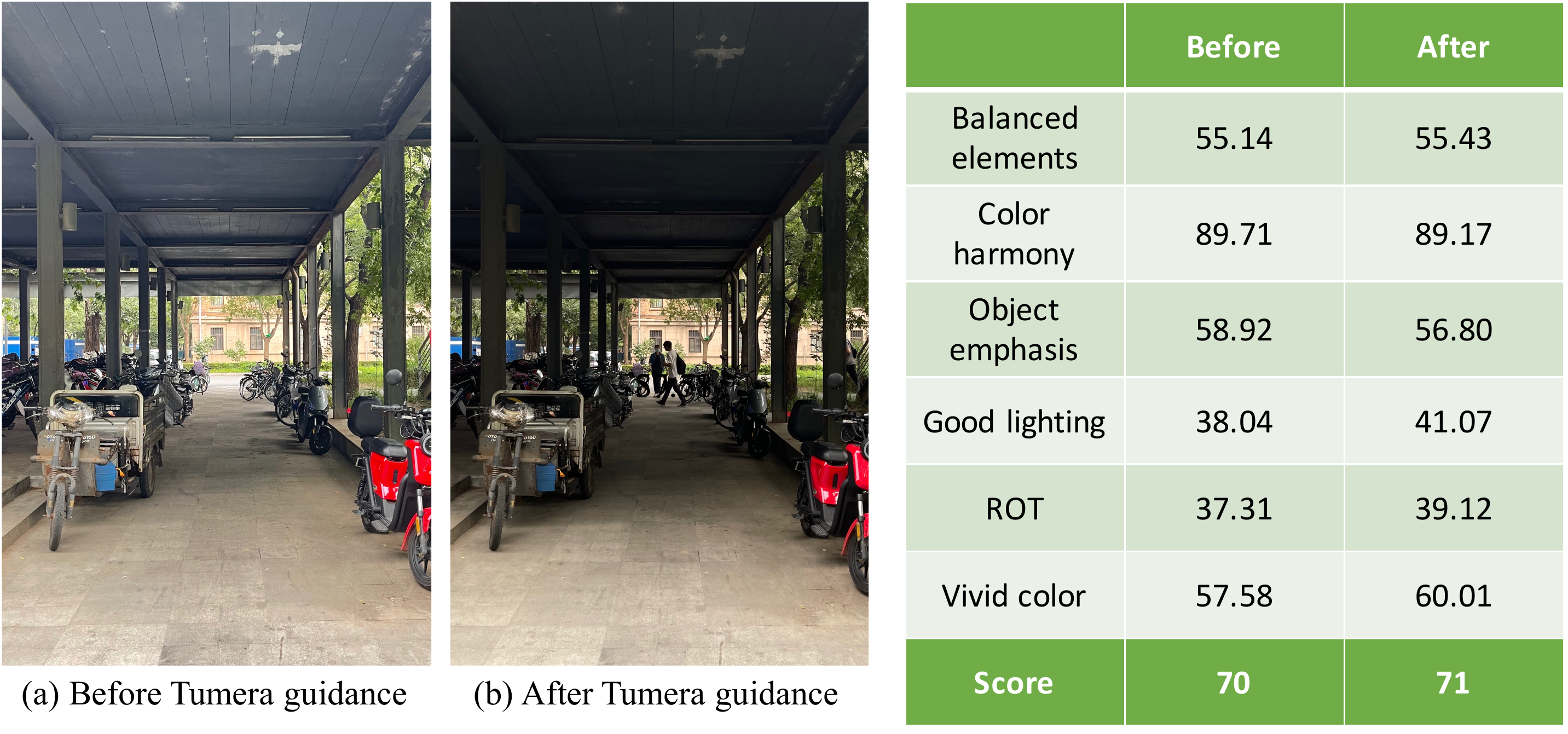}
    \caption{Scores of a pair of photos taken before and after using Tumera.}
    \label{fig:score_comparison}
\end{figure}
The experimental results (Table~\ref{tab:diff}) show that the average score of the tester’s photos increases by 13 points after using Tumera, and 83\% of the works shoot after the photography guidance are better than the unsupervised works. The largest difference in score is 33 points, and 53.3\% of the score differences are around 10 points. These results demonstrate that Tumera can help photography beginners improve their shooting skills. In Fig.~\ref{fig:score_comparison}, we show a pair of photos and their scores.

\begin{table}[ht]
    \centering
    \caption{Comparison of scores before and after the guidance of Tumera system.}\label{tab:diff}
    \begin{tabular}{p{1.5cm}p{0.8cm}p{0.8cm}p{0.8cm}p{0.8cm}p{0.8cm}p{0.8cm}p{0.8cm}p{0.8cm}p{0.8cm}p{0.8cm}}
    \hline
         Group 1 & 1 & 2 & 3 & 4 & 5 & 6 & 7 & 8 & 9 & 10 \\
    \hline
         Before  & 63 & 55 & 71 & 68 & 45 & 66 & 72 & 65 & 61 & 68 \\
         After   & 72 & 61 & 63 & 93 & 56 & 68 & 65 & 83 & 94 & 84 \\
         Diff.   & 9  & 6  & -8 & 25 & 11 & 2  & -7 & 18 & 33 & 16 \\
    \hline
    \end{tabular}
    
\vspace{1.5em}
    \begin{tabular}{p{1.5cm}p{0.8cm}p{0.8cm}p{0.8cm}p{0.8cm}p{0.8cm}p{0.8cm}p{0.8cm}p{0.8cm}p{0.8cm}p{0.8cm}}
    \hline
         Group 2 & 21 & 22 & 23 & 24 & 25 & 26 & 27 & 28 & 29 & 30 \\
    \hline
         Before  & 75 & 52 & 39 & 67 & 76 & 89 & 70 & 54 & 78 & 67 \\
         After   & 89 & 67 & 52 & 84 & 71 & 76 & 71 & 83 & 89 & 73 \\
         Diff.   & 14 & 15 & 13 & 17 & -5 & -13& 1  & 29 & 11 &  6 \\
    \hline
    \end{tabular}
    
\vspace{1.5em}
    \begin{tabular}{p{1.5cm}p{0.8cm}p{0.8cm}p{0.8cm}p{0.8cm}p{0.8cm}p{0.8cm}p{0.8cm}p{0.8cm}p{0.8cm}p{0.8cm}}
    \hline
         Group 3 & 11 & 12 & 13 & 14 & 15 & 16 & 17 & 18 & 19 & 20 \\
    \hline
         Before  & 74 & 30 & 66 & 81 & 78 & 53 & 61 & 54 & 35 & 63 \\
         After   & 63 & 48 & 81 & 79 & 91 & 65 & 61 & 73 & 41 & 67 \\
         Diff.   & -11& 18 & 15 & -2 & 13 & 12 & 0  & 19 &  6 & 4 \\
    \hline
    \end{tabular}
    
\end{table}

\emph{Photography professionals make better choices for 30 sets of photos.} After the 30 sets of photos are shot, they are handed over to three photography professionals without being told which picture is taken after Tumera's guidance. Professionals are requested to select the better photo for each of 30 photo pairs. Table~\ref{tab:consistency} shows that 21, 19, and 25 pictures selected by the advertising photographer, the photography graduate student, and the college photography teacher are taken under the guidance of Tumera. The coincidence rates are: 70\%, 63.3\%, 83.3\%. The overall coincidence rate is 72.2\%, indicating that the Tumera system has a high degree of objectivity, professionalism, and stability in aesthetic evaluation. These results also further consolidate that Tumera can be used to guide photography beginners in photography learning and aesthetic evaluation. Photography professionals also give reasons for the choice, and propose improvements for some of the six characteristic dimensions. 

\begin{table}[ht]
    \centering
    \caption{Whether aesthetic evaluations given by Tumera and professionals are consistent.}\label{tab:consistency}
    \begin{tabular}{p{2.5cm}p{0.5cm}p{0.5cm}p{0.5cm}p{0.5cm}p{0.5cm}p{0.5cm}p{0.5cm}p{0.5cm}p{0.5cm}p{0.5cm}}
    \hline
         Group 1 & 1 & 2 & 3 & 4 & 5 & 6 & 7 & 8 & 9 & 10 \\
    \hline
         Ad photog  & \checkmark & \checkmark &  & \checkmark & \checkmark &  & \checkmark &  & \checkmark & \checkmark \\
         Photog graduate & \checkmark & \checkmark &  & \checkmark &  & \checkmark &  &  & \checkmark & \checkmark \\
         Photog teacher  & \checkmark  & \checkmark  & \checkmark & \checkmark & \checkmark & \checkmark  & \checkmark & \checkmark & \checkmark & \checkmark \\
    \hline
    \end{tabular}
    
\vspace{1.5em}
    \begin{tabular}{p{2.5cm}p{0.5cm}p{0.5cm}p{0.5cm}p{0.5cm}p{0.5cm}p{0.5cm}p{0.5cm}p{0.5cm}p{0.5cm}p{0.5cm}}
    \hline
         Group 2 & 11 & 12 & 13 & 14 & 15 & 16 & 17 & 18 & 19 & 20 \\
    \hline
         Ad photog  & \checkmark & \checkmark &  &  & \checkmark & \checkmark & \checkmark &  &  & \checkmark \\
         Photog graduate   & \checkmark & \checkmark &  &  & \checkmark &  & \checkmark & \checkmark &  & \checkmark \\
         Photog teacher   & \checkmark & \checkmark &  & \checkmark & \checkmark & \checkmark & \checkmark  &  &   & \checkmark \\
    \hline
    \end{tabular}
    
\vspace{1.5em}
    \begin{tabular}{p{2.5cm}p{0.5cm}p{0.5cm}p{0.5cm}p{0.5cm}p{0.5cm}p{0.5cm}p{0.5cm}p{0.5cm}p{0.5cm}p{0.5cm}}
    \hline
         Group 3 & 21 & 22 & 23 & 24 & 25 & 26 & 27 & 28 & 29 & 30 \\
    \hline
         Ad photog  & \checkmark & \checkmark &  & \checkmark & \checkmark & \checkmark & \checkmark & \checkmark & \checkmark &  \\
         Photog graduate   & \checkmark & \checkmark &  & \checkmark & \checkmark & \checkmark & \checkmark & \checkmark & \checkmark &  \\
         Photog teacher   & \checkmark & \checkmark & \checkmark & \checkmark & \checkmark & \checkmark&   & \checkmark &  & \checkmark  \\
    \hline
    \end{tabular}
\end{table}

%% file: 6-Conclusion.tex
\section{Conclusion}
In this paper, we introduce an interactive assistant system, Tumera, that helps photography beginners learn to take photos. Tumera achieves this by modeling the learning process in three different steps: perception, action, and feedback. Tumera provides real-time feedback during shooting which prompts users to change composition and brightness of photos. Aesthetic evaluation is given after photos are taken, enabling users to know whether their photographic works are of high quality. Users can upload their works to the Tumera community, where they can share opinions about photos. Tumera also score the uploaded photos, rank them, and give recommendations to users accordingly. Hopefully, these interactions will motive users to continuously improve their photography skills. Tumera's future plan intends to add more feature dimensions into the model to better help photography beginners, to provide more guidance at the action level, and to give more comprehensive feedback at the feedback level.